\documentclass[prl,twocolumn,amsmath,amssymb,superscriptaddress,nofootinbib]{revtex4-2}
\usepackage{graphicx}
\usepackage{amsmath}
\usepackage{yhmath}
\usepackage{bm}
\usepackage{xcolor}
\usepackage{epstopdf}
\usepackage{subfigure}
\usepackage{hyperref}
\usepackage[T1]{fontenc}
\usepackage{lmodern}

\begin{document}

\title{The fate of disorder in twisted bilayer graphene near the magic angle}

\author{Zhe Hou}
\email{zhe.hou@nnu.edu.cn}
\affiliation{School of Physics and Technology, Nanjing Normal University, Nanjing 210023, China}
\author{Hailong Li}
\affiliation{Interdisciplinary Center for Theoretical Physics and Information Sciences (ICTPIS), Fudan University, Shanghai
200433, China}
\author{Qing Yan}
\affiliation{Interdisciplinary Center for Theoretical Physics and Information Sciences (ICTPIS), Fudan University, Shanghai 200433, China}
\author{Yu-Hang Li}
\affiliation{School of Physics, Nankai University,Tianjin 300071, China}
\affiliation{State Key Laboratory of the Surface Physics, Fudan University, Shanghai 200433, China}
\author{Hua Jiang}
\email{jianghuaphy@fudan.edu.cn}
\affiliation{Interdisciplinary Center for Theoretical Physics and Information Sciences (ICTPIS), Fudan University, Shanghai 200433, China}

\begin{abstract}
{In disordered lattices, itinerant electrons typically undergo Anderson localization due to random phase interference, which suppresses their motion. By contrast, in flat-band systems where electrons are intrinsically localized owing to their vanishing group velocity, the role of disorder remains elusive. Twisted bilayer graphene (TBG) at the magic angle $\sim 1.1^\circ$ provides a representative flat-band platform to investigate this problem. Here, we perform an atomistic tight-binding quantum transport calculation on the interplay between disorder and flat-bands in TBG devices. This non-phenomenological approach provides direct evidence that moderate disorder enhances conductance, whereas stronger disorder restores localization, revealing a disorder-driven delocalization-to-localization transport behavior. The underlying physical mechanism is understood by an effective inter-moir{\'e} tunneling strength via spectral flow analysis of a disordered TBG cylinder. Moreover, by comparing magic-angle and large-angle TBG, we demonstrate qualitatively distinct disorder responses tied to the presence of flat-bands. Our quantitative results highlight the unconventional role of disorder in flat-band moir{\'e} materials and offer insights into the observation of the fractional quantum anomalous Hall effect in disordered moir{\'e} systems.} 
\end{abstract}

\maketitle

\emph{Introduction.} 
Disorder is widely recognized as detrimental to electronic transport in condensed matter systems. A central example is Anderson localization (AL), where itinerant particles in regular lattices become completely localized due to random on-site disorder~\cite{Anderson1958Absence}. On the other hand, localization can also occur in disorder-free systems with flat energy bands arising from specific lattice geometries, such as the two-dimensional (2D) Lieb~\cite{Vicencio2015Observation, Mukherjee2015Observation, Taie2015Coherent, Silva2014Experimental}, kagome lattices~\cite{Takeda2004Flat, Bergman2008Band, Endo2010Tight, Masumoto2012Exciton, Jo2012Ultracold, Zong2016Observation}, the one-dimensional (1D) Tasaki lattice~\cite{Tasaki1992Ferromagnetism, Mielke1993Ferromagnetism, Liu2019Flat}, and Aharonov-Bohm (AB) cages~\cite{Vidal1998Aharonov, Mukherjee2018Experimental, Chen2025Interaction}. This phenomenon, termed flat-band localization (FBL) or geometric localization, results from destructive interference in particle hopping. Recently, the interplay between disorder effect and flat-bands has attracted growing interest~\cite{Chalker2010Anderson, Bodyfelt2014Flatbands, Flach2014Detangling, LeyKam2017Localization, Roy2020Interplay, Gligoric2020Influence, Cadez2021Metal, Zuo2024Topological, Zeng2024Transition, Mao2024Transition, Goda2006Inverse, Longhi2021Inverse, Li2022Aharonov, Wang2022Observation, Zhang2023NonAbelian}, uncovering a variety of intriguing phenomena such as unconventional scaling of localization lengths~\cite{LeyKam2017Localization}, critical states that are neither Anderson-localized nor spatially extended~\cite{Chalker2010Anderson}, a transition from FBL to AL~\cite{Zeng2024Transition, Mao2024Transition}, and even an inverse Anderson transition~\cite{Goda2006Inverse, Longhi2021Inverse, Li2022Aharonov, Wang2022Observation, Zhang2023NonAbelian}.

In real condensed-matter systems, flat-bands with the Fermi level nearby are exceedingly rare, typically realized in quantum Hall systems \cite{Klitzing1980New, Laughlin1981Quantized, Halperin1982Quantized, Thouless1982Quantized} and twisted moir{\'e} structures \cite{Santos2007Graphene, Bistritzer2010Morie, Santos2012ContinuumModel, Morell2010Flat, Wong2020Electronic, Andrei2020Graphene, Chen2020Configurable, Wang2020Correlated, Kennes2021Moire, YCao2018Unconventional, Yankowitz2019Tuning, Christos2022Correlated, Xia2025Superconductivity, Guo2024Superconductivity, Xia2025Superconductivity, Guo2024Superconductivity, YCao2018Correlated, Stepanov2020Untying, Cao2020Tunable, Chen2021Electrically, Sharpe2019Emergent, Lin2022Spin, Zhang2020Correlation, Saito2021Hofstadter, Serlin2020Intrinsic, Wu2021Chern, Cai2023Signatures, Park2023Observation, Xu2023Observation, Kang2024Evidence}. The latter have become a major research focus, hosting diverse correlated states such as unconventional superconductivity \cite{YCao2018Unconventional, Yankowitz2019Tuning, Christos2022Correlated, Xia2025Superconductivity, Guo2024Superconductivity, Xia2025Superconductivity, Guo2024Superconductivity}, correlated insulators \cite{YCao2018Correlated, Stepanov2020Untying, Cao2020Tunable, Chen2021Electrically}, ferromagnetism \cite{Sharpe2019Emergent, Lin2022Spin, Zhang2020Correlation, Saito2021Hofstadter}, quantum anomalous Hall (QAH) states \cite{Serlin2020Intrinsic, Wu2021Chern}, fractional QAH states \cite{Cai2023Signatures, Park2023Observation, Xu2023Observation}, and fractional quantum spin Hall phases \cite{Kang2024Evidence}. Most theoretical descriptions of the above phenomena assume clean bands and overlook disorder, however, prior studies of Landau flat-bands have shown that disorder can strongly influence the formation of fractional quantum Hall and related quantum states \cite{Tsui1982Two, Paalanen1984Disorder, MacDonald1986Disorder, Wan2005Mobility, Wan2011Impact, Deng2014Disorder, Zhu2019Disorder, Williams1991Conduction, Li2000Microwave}, an effect that may be equally crucial in moir{\'e} flat-bands. This challenge is amplified by the ubiquitous strong disorder in moir{\'e} systems—point defects, impurities, and twist-angle inhomogeneity \cite{Guo2021Moire, Sari2023Analysis, Zhao2023Excitons, Fang2023Localization, Jong2022Imaging, Cruz2021High, Uri2020Mapping}—as exemplified by the fractional QAH effect recently observed in twisted TMDs \cite{Cai2023Signatures, Park2023Observation, Xu2023Observation}, where intrinsic short-range impurities are believed to be ubiquitous. To capture such effects, atomistic modeling is necessary, but small twist angles lead to supercells containing thousands of atoms \cite{Koshino2018Maximinally, Angeli2021Gamma, Yang2024Evolution}, making transport calculations highly challenging. To date, wave-function diffusion simulations have mainly offered indirect insights into this problem \cite{Guerrero2025Disorder}, while direct quantitative conductance calculations are still lacking.

\begin{figure}[ttt]
\includegraphics[width=8.5cm, clip=]{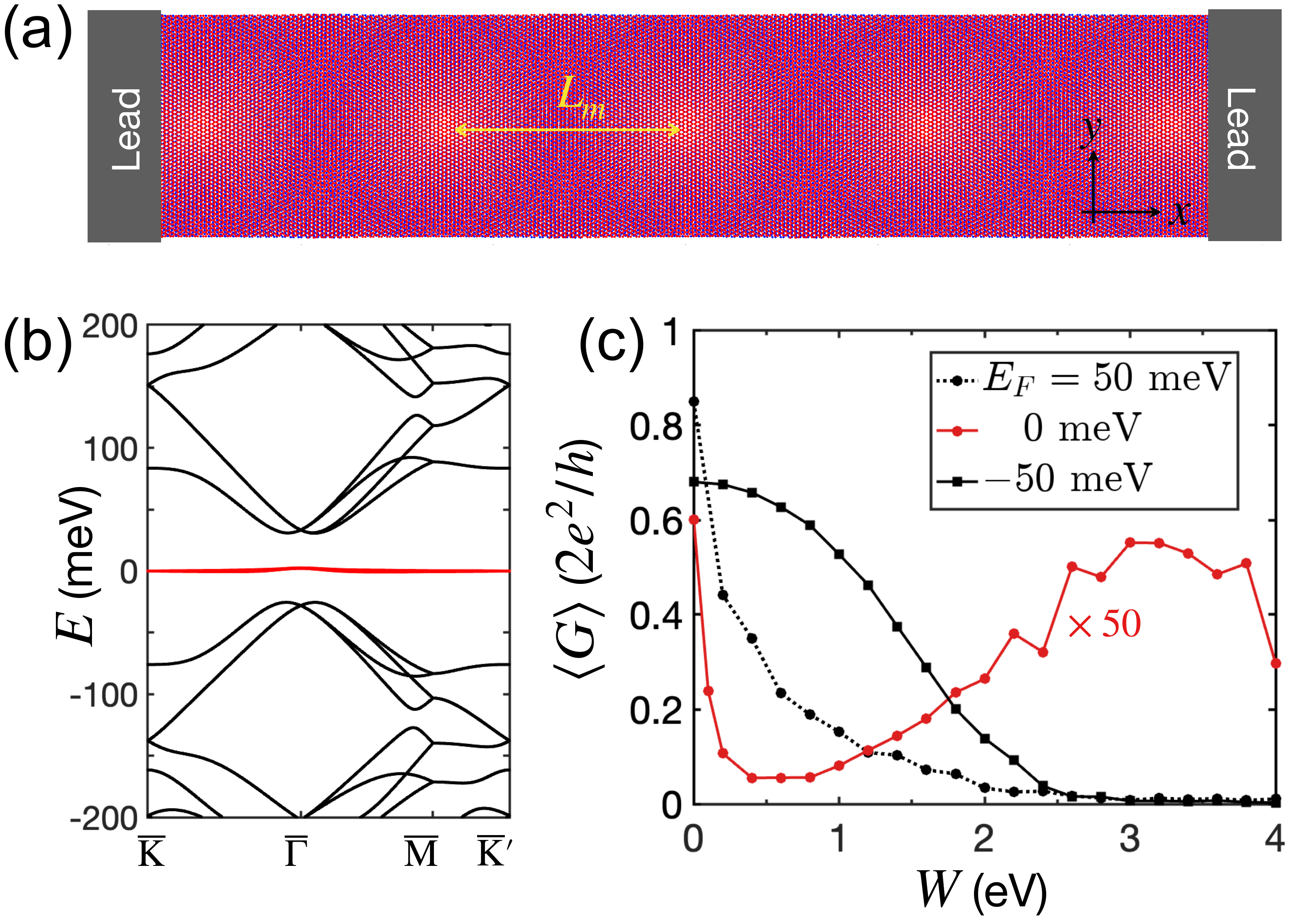}
\caption{(a) Schematic diagram of the two-terminal transport device with a rectangular TBG flake sandwiched between two metallic leads (shown in dark). The moir{\'e} length is denoted as $L_m$. (b) Band structure of the TBG along $\rm \overline{K}-\overline{\Gamma}-\overline{M}-\overline{K}'$ in the moir{\'e} Brillouin zone. Here a commensurate twist angle $\theta \approx 1.12^\circ$ is chosen, characterized by two integer numbers $m=29, n=30$. (c) Ensemble-averaged conductance $\langle G \rangle $ as a function of the disorder strength $W$. Here for the red (blue) curve, the Fermi energy $E_F$ is set at the dispersive conduction (valence) band, as marked by the red (blue) dashed line in the band structure of (b). For the black curve, $E_F$ is set at the flat band ($E_F=0$). For the sake of comparison, $\langle G \rangle $ at $E_F=0$ has been amplified by 50 times. The transport parameters are: $N=55, L=261$, $\eta_c=0.5$ meV. $\langle G \rangle $ is averaged over 100 configurations for the dispersive bands, and over 500 configurations for the flat-bands. }
\label{fig.Setup_Band}
\end{figure}

In this work, we propose a practical protocol to study the interplay between disorder and flat-bands by constructing a quasi-one-dimensional (quasi-1D) twisted bilayer graphene (TBG) nanoribbon model [Fig.~\ref{fig.Setup_Band}(a)], enabling fully quantum atomistic conductance calculations. As disorder strength increases, the conductance of flat-band electrons first decreases under weak disorder and then shows an anomalous enhancement at moderate disorder [Fig.~\ref{fig.Setup_Band}(c)], indicating that initially localized flat-band electrons become delocalized and gain mobility. In contrast, electrons in dispersive bands at higher energies or at the charge neutrality point (CNP) with larger twist angles exhibit rapid conductance suppression, restoring the conventional AL. These results establish disorder-enhanced transport as a distinctive feature of flat-band electrons. Notably, this work delivers the first fully quantum, quantitative transport calculations on a realistic atomic twisted lattice—unlike previous indirect approaches~\cite{Guerrero2025Disorder} or low-energy continuum models~\cite{Koshino2018Maximinally, Kang2018Symmetry, Song2022Magic}, which treat the moir{\'e} supercell as a whole and cannot capture short-range impurities. The platform also offers a robust route to explore disorder–flat-band interactions in moir{\'e} systems and may help elucidate the fractional QAH effect observed in twisted bilayer MoTe$_2$~\cite{Cai2023Signatures, Park2023Observation, Xu2023Observation}, where intrinsic impurities are ubiquitous.

\emph{Model and methods.}
As shown in Fig.~\ref{fig.Setup_Band} (a), we consider a rectangular TBG nanoflake with length $\mathcal{D}_L$ and width $\mathcal{D}_W$ mediated between two metallic leads. An AA-stacking QD-like array is formed inside the nanoflake, at which the flat-band electrons mainly reside~\cite{Laissardiere2010Localization, Laissardiere2012NumericalStudies}. The TBG is obtained by twisting the top (bottom) layer with angle $\pm \theta/2$. Cutoff boundaries are adopted along the $y$-direction, as can be realized by physical etching~\cite{PalauTwistHetero2018, MahapatraQHTBG2022} or electrical gating~\cite{DeschenesTBGEdgeStates2022, VriesTBGJosephson2021}. The twist angle $\theta$ is chosen to be commensurate, characterized by two integer numbers $(m, n)$ obeying the relation $\cos{\theta} = {m^2 + 4 mn + n^2}/{[2(m^2 + mn + n^2)]}$~\cite{Laissardiere2010Localization, Laissardiere2012NumericalStudies}. In this case, the central QD-like array has a strict moir{\'e} period $L_m = {a_0 }/{[2 \sin{(\theta/2)}] }$ considering an infinitely-long nanoribbon, with $a_0 = 2.46 {\rm \AA}$ the lattice constant of monolayer graphene. Below we use two integer numbers $L$ and $N$ to characterize the size of the central nanoflake, satisfying $\mathcal{D}_L = \sqrt{3}L a_0 $ and $ \mathcal{D}_W = N a_0$. 

To conduct the atomistic simulation, we use the following tight-binding Hamiltonian to describe the central TBG region~\cite{Laissardiere2010Localization, Laissardiere2012NumericalStudies}:
\begin{align}
H= \sum_i |i \rangle \varepsilon_i \langle i | + \sum_{ i,j } |i \rangle t_{ij} \langle j| ,
\end{align} 
where $| i \rangle$ denotes the $p_z$ orbital of a carbon atom at position ${\bm r}_i$, and $\varepsilon_i$ is the on-site energy. We consider the Anderson disorder in the TBG region, where $\varepsilon_i$ is uniformly distributed within $[-W/2, W/2]$ with $W$ being the disorder strength. The coupling integral $t_{ij}$ obeys the Slater-Koster relation~\cite{Slater1954HoppingFunction}:
\begin{align}
t_{ij} =  \chi^2 V_{pp \sigma} (r_{ij}) + (1- \chi ^2)V_{pp \pi}(r_{ij}) ,
\label{eq: t_ij}
\end{align}
where $\chi$ is the direction cosine along the $z$-direction: $\chi \equiv {{\bf r}_{ij} \cdot {\bf e}_z}/{r_{ij}}$ with $r_{ij} = | {\bf r}_j - {\bf r}_i|$ the atomic distance and ${\bf e}_z$ the $z$-direction unit vector. The $pp \sigma$ and $pp \pi$ types of hopping integral reads $V_{pp \sigma} (r_{ij}) =  \gamma_1 e^{ (a_I -r_{ij}) q_{\sigma}/ {a_I} }$, and $V_{pp \pi} (r_{ij}) = - \gamma_0 e^{  (a-r_{ij}) q_{\pi}/ {a}}$, where $a_I = 3.349 {\rm \AA}$ is the layer distance in the AB-stacking region (we consider lattice corrugation effect) and $a=1.418 {\rm \AA}$ is the intra-layer carbon-carbon atomic distance. The parameters are set to $\gamma_0 = 2.7$ eV, $\gamma_1 = 0.48$ eV, and $q_{\pi}/ {a} = q_{\sigma}/ {a_I} = 2.218$ ${\rm \AA}^{-1}$ to fit the DFT calculations~\cite{Laissardiere2010Localization, Laissardiere2012NumericalStudies}. Since the hopping integral decays exponentially with distance $r_{ij}$, a cutoff range of $3a$ has been adopted for the in-plane hopping. 

It has been shown that the lattice corrugation effect, i.e. the non-equivalence of interlayer distances between AB/AA-stacking regions, is important in accounting for the large band gap between the flat-bands and the remote dispersive bands~\cite{Koshino2018Maximinally, Liu2019Pseudo}. To include this effect, we use the following $\bf r$-dependent interlayer distance: 
\begin{align}
d({\bf r}) = d_0 + 2 d_1 \sum_{j=1}^3  \cos{[{\bf b}_j \cdot \delta({\bf r})]},
\end{align}
where $\delta({\bf r}) = [R(\theta/2) - R(-\theta/2)] \cdot {\bf r} $ is the in-plane shift with $R(\theta)$ the rotation matrix. ${\bf b}_1 = [{2 \pi} /{( \sqrt{3} a_0} ), {2 \pi}/{a_0}] $, ${\bf b}_2 = [-{4 \pi } /{(\sqrt{3} a_0)}, 0] $, and ${\bf b}_3 = -{\bf b}_1 - {\bf b}_2$ are the three reciprocal lattice vectors of the untwisted monolayer graphene. In the numerical calculations, we take $d_0 = 3.433 \AA$ and $d_1 = 0.0278 \AA$. 

The two-terminal differential conductance $G$ through the disordered TBG at zero temperature can be calculated using the non-equilibrium Green's function method~\cite{Meir1992LandauerFormula, Jauho1994TransportResonant, Hou2024TBG, Hou2024Arrays}:
 \begin{align}
G(E_F) = \frac{2e^2}{h} {\rm Tr} [ {\bf \Gamma}_L {\bf G}^r_C {\bf \Gamma}_R {\bf G}^a_C],
\end{align}
where $E_F$ is the Fermi energy (defined relative to the CNP of a TBG nanoribbon~\cite{CNPDefinition}), ${\bf \Gamma}_{L(R)}$ are the linewidth functions for leads, and ${\bf G}^{r(a)}_C(E_F)$ is the retarded (advanced) Green's function of the central region, expressed by: ${\bf G}^{r(a)}_C(E_F) = [(E_F \pm i \eta_c){\bf I} - {\bf H}_C - {\bf \Sigma}^r_L - {\bf \Sigma}^r_R ] ^{-1}$, with ${\bf H}_C$ the Hamiltonian matrix of the central region and $\eta_c$ the imaginary line-width function~\cite{Li2024Emergent}. Since the band-width of the metallic leads is orders of magnitude larger than the TBG, we use the wide-band approximation~\cite{Bahamon2024Chirality, Bahamon2020Emergent} ${\bf \Gamma}_{L(R)} = -i \gamma_0 {\bf I}_{L(R)}$ where ${\bf I}_{L(R)}$ is the identity matrix with dimension determined by the number of coupled carbon atoms with the leads.

\begin{figure}[ttt]
\includegraphics[width=8.4cm, clip=]{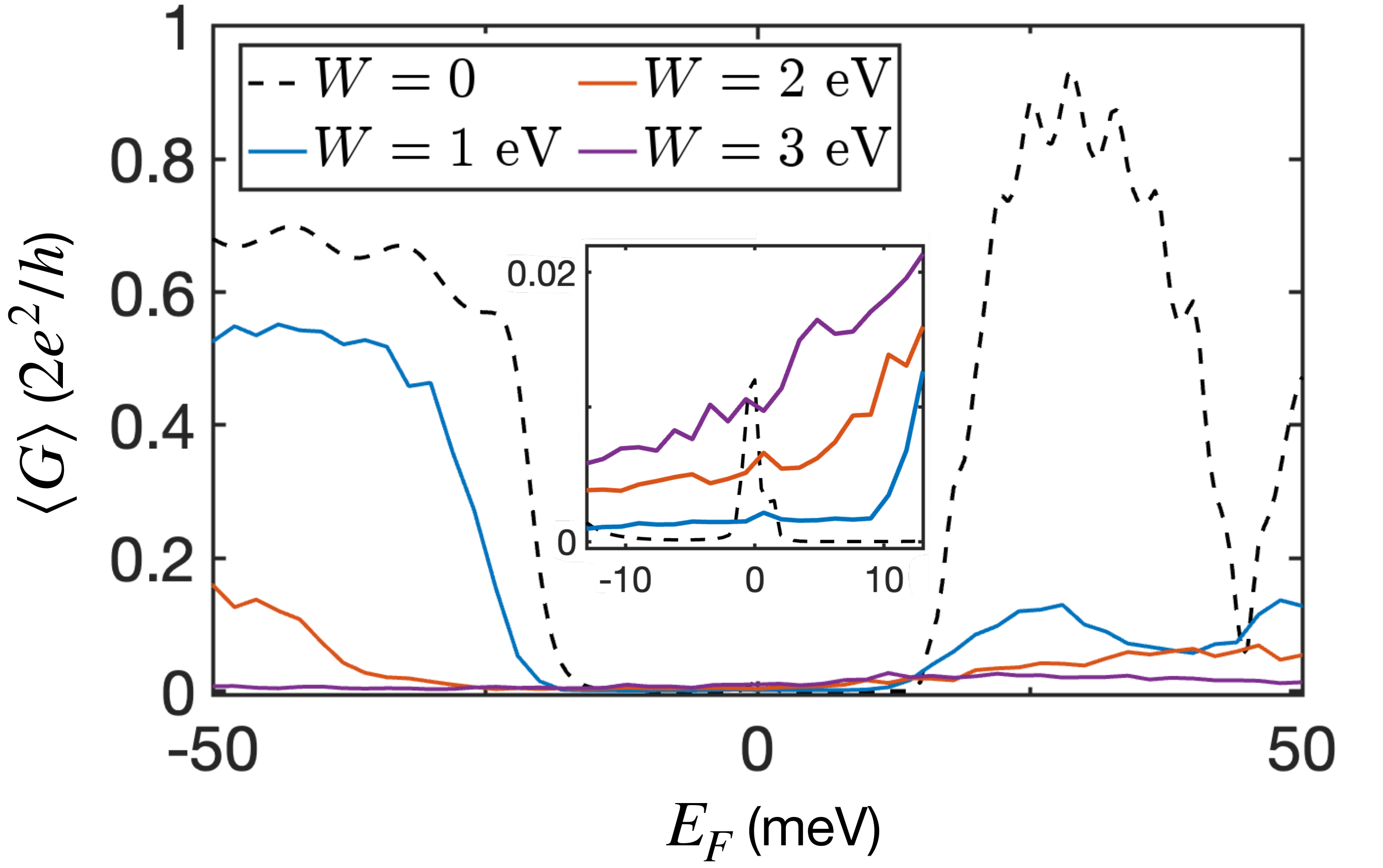}
\caption{ $\langle G \rangle$ as a function of $E_F$ at different disorder strength $W$. Inset: zoom-in plot of the curves near the CNP. Curves in the inset are averaged over 500 configurations while others are averaged over 100 configurations. The parameters are: $N=55, L=261$, $\eta_c=0.5$ meV.  }
\label{fig.G_ScanEf}
\end{figure}

The flat-bands appear at the magic angle $\sim 1.1^{\circ}$, we thus set $m=29$ and $n=30$, in which case the twist angle is $\theta \approx  1.12^{\circ}$ (we have checked other values around it and obtained the same conclusion, see S.M.~\cite{SM} for details). In Fig.~\ref{fig.Setup_Band}(b), we plot the band structure of a 2D TBG along $\rm \overline{\Gamma} - \overline{K} - \overline{M} $ in the moir{\'e} Brillouin zone. Two flat-bands, which are spin-valley fourfold degenerate, emerge at the CNP. A band gap of about 25 meV appears due to the lattice corrugation, separating the flat-bands and remote dispersive bands. Note that the flat-bands have a tiny band-width of $\sim4$meV. When the Fermi energy lies at the CNP, coinciding with the flat-bands, electrons are completely confined within the AA-stacking region inside the QD because of the strong inter-layer coupling. However, the AB(BA)-stacking regions, where inter-layer coupling is weak, are almost empty. We thus consider a TBG nanoflake terminating at AB(BA)-stacking spots along the $y$-direction by setting $N=55$. In this case, the compact QD array is formed by one row of repeated AA-stacking regions as illustrated in Fig.~\ref{fig.Setup_Band}(a).  

\emph{Disorder enhanced conductance.}  Figure~\ref{fig.Setup_Band}(c) shows the disorder-averaged two-terminal conductance $\langle G \rangle $ as a function of the disorder strength $W$ for three different Fermi energies $E_F=-50$ meV, $0$ meV, and $50$ meV. When $E_F=\pm50$  meV, the Fermi energy intersects with the dispersive bands and the system is thus a metal. Consequently, the averaged conductance $\langle G \rangle$ decreases monotonically as $W$ increases, and finally approaches zero when $W>3$ eV, exhibiting typical AL behavior. Moreover, the conductance in the dispersive conduction band decreases more sharply than in the valence bands owing to the particle-hole asymmetry. Notably, when the Fermi energy coincides with the flat-bands at $E_F=0$ meV, as the disorder strength increases, the averaged conductance $\langle G \rangle$ first decreases sharply under weak disorder ($W<0.4$ eV), then increases anomalously under moderate disorder ($0.4$ eV$<W<3$ eV), featuring delocalization behavior. The conductance decreases again under strong disorder with strength $W>3$ eV, which indicates a crossover from delocalization to AL. Our result provides a global picture for the conductance evolution of the flat-band versus the disorder strength, different from Ref.~\cite{Guerrero2025Disorder} that only three discrete disorder strengths are considered.

\begin{figure}[ttt]
\includegraphics[width=7.8cm, clip=]{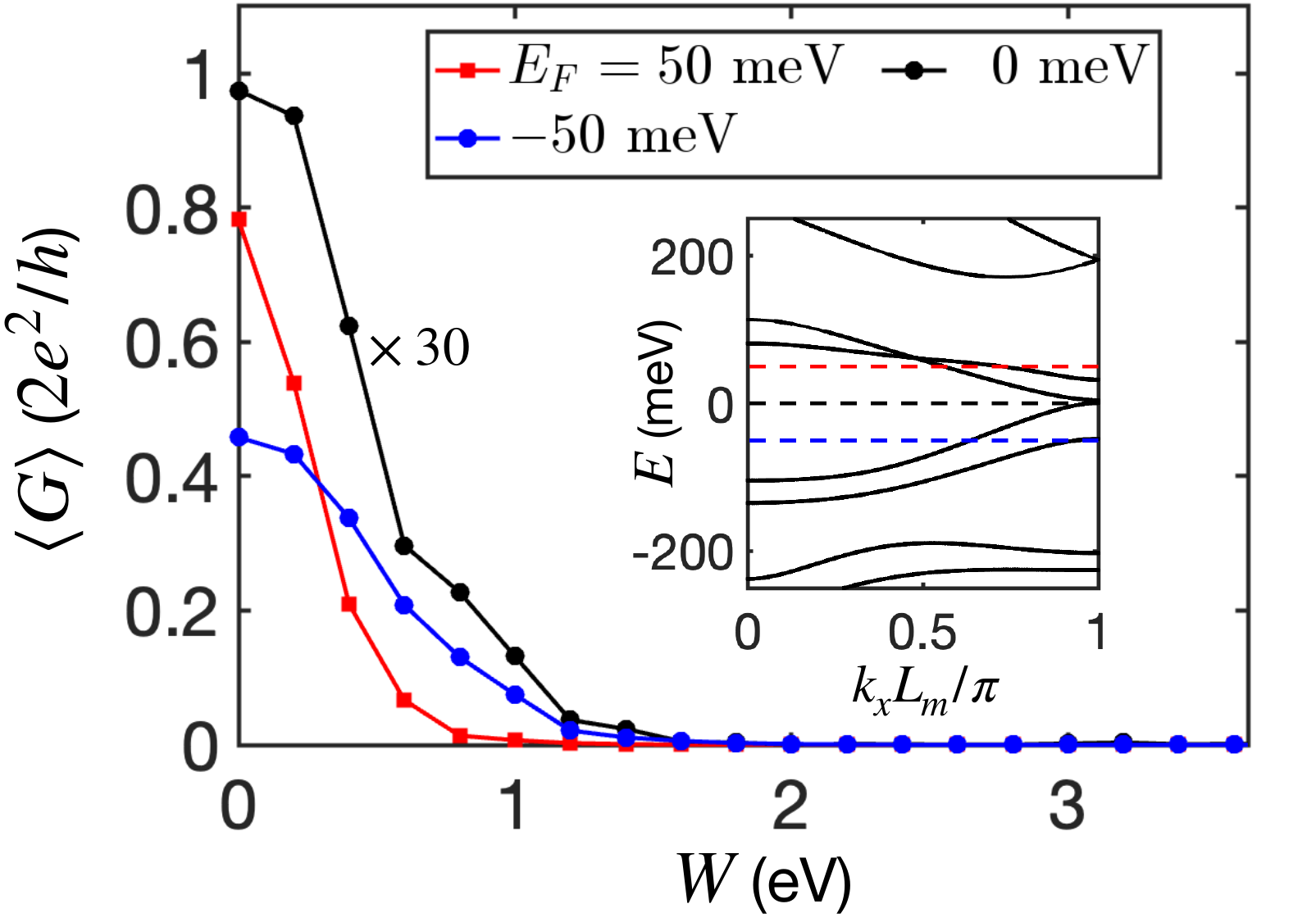}
\caption{Numerical results for twist angle $\theta \approx 3.15^{\circ}$, characterized by two commensurate numbers $m=10$ and $n=11$. $\langle G \rangle $ versus $W$ at different $E_F$. $\langle G \rangle $ at $E_F=0$ is amplified by 30 times.  The inset shows the band structure of a quasi-1D TBG nanoribbon with flat cutoff boundaries with width $N=23$. }
\label{fig.m10n11}
\end{figure}

\begin{figure*}
\includegraphics[width=18cm, clip=]{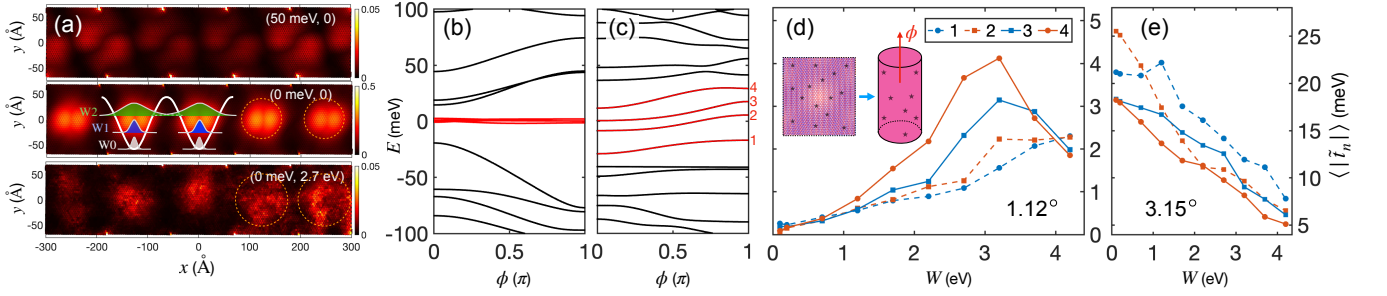}
\caption{(a) Distribution of LDOSs on the bottom layer of the central TBG flake shown in Fig. 1(a). Here the energy and disorder strength are labeled on the top right part inside each panel as $(E, W)$. The inset in the middle panel illustrates the broadening of wavefunctions inside the TBG nanoflake for different values of disorder strength $W$. The white dashed circles indicate the region of the localized wavefunctions. (b) and (c): Spectral flow of a periodic clean and disordered TBG ribbon as shown in the inset of (d). For (c) one disorder configuration with disorder strength $W=2.7$ eV is chosen, and the lowest energy bands close to the CNP are labeled by number 1 to 4 in energy-ascending order. (d) The ensemble-averaged effective inter-QD tunneling strength $\langle |\tilde{t}_n| \rangle$ of the disordered moir{\'e} supercell ribbon (shown in the inset), as a function of $W$ for the four lowest energy bands marked in (c). (e) $\langle |\tilde{t}_n| \rangle$ versus $W$ at twist angle $\theta \approx 3.15^{\circ}$. To obtain $\langle |\tilde{t}_n| \rangle$ 21 points of $\phi$ are adopted. The other parameters are: $N=55$, $\theta \approx 1.12^{\circ}$ and $\eta_c=0.5$ meV for (a-d), and  $N=23$ for (e). For each $W$, 100 configurations are considered.   }
\label{fig.LDOS_SpectralFlow}
\end{figure*}

In Fig.~\ref{fig.G_ScanEf}, we present a detailed Fermi energy ($E_F$)-dependence of $\langle G \rangle$ under disorder strengths: $W=0, 1, 2$ and 3 eV. For $W=0$, i.e., disorder-free system, we identify two different energy regions from the conductance values: the dispersive-bands region, where the conductance has high values, and the flat-bands region, where conductance is close to zero. Note that the conductance at $E_F = 0$ is not zero (see the conductance peak in the inset of Fig.~\ref{fig.G_ScanEf}) since the flat-bands still have a tiny but nonzero band-width ($\sim$ 4 meV), distinct from the complete flat-bands in the Tasaki lattice \cite{Tasaki1992Ferromagnetism, Mielke1993Ferromagnetism, Liu2019Flat, Zeng2024Transition, Mao2024Transition} or the Aharonov-Bohm cage \cite{Vidal1998Aharonov, Mukherjee2018Experimental, Chen2025Interaction, Longhi2021Inverse, Li2022Aharonov}. When $W=1$ eV, both conductance in the dispersive-bands region and that at $E_F=0$ decrease. However, as $W$ further increases, conductance in the flat-bands region all rise, while conductance in the dispersive region decreases monotonically with $W$. This means that the disorder-enhanced conductance is a unique property of the flat-band region.

We analyze the mechanism of flat-band conductance under different disorder strengths [see Fig.~\ref{fig.Setup_Band} (c)]. The twist-induced flat-bands, with a band-width of $\sim$ 4 meV, are well separated from dispersive bands [Fig.~\ref{fig.Setup_Band} (b)]. Thus, even $W=0.1$ eV—about 30 times larger than the band-width, acts as strong disorder, leading to pronounced localization. This explains the AL behavior of $\langle G \rangle$ for $W<0.4$ eV.
At intermediate disorder ($0.4 \leq W \leq 3 $ eV), however, $\langle G \rangle$ exhibits an anomalous enhancement, deviating from AL and indicating an unexpected transport mechanism. Importantly, this unexpected mechanism should be related to the unique flat-band at the magic-angle. Once the energy is tuned into the remote dispersive bands, or the twist angle becomes larger, the anomalous behavior disappears and only AL happens. For example, Fig.~\ref{fig.m10n11} exemplifies the conductance at $\theta \approx 3.15^{\circ}$, where $\langle G \rangle$ decreases rapidly with $W$ at all Fermi energies.

\emph{Mechanism of disorder-enhanced conductance at the magic angle.} 
The unexpected transport mechanism can be understood from the real-space wavefunctions corresponding to the flat-bands, which initially localize inside the QD, and then spread out under scattering of disorder. In Fig.~\ref{fig.LDOS_SpectralFlow}(a), we plot the local density of states (LDOSs) $\rho({\bf r}_i)$ distributed inside the central TBG region, as calculated by $\rho(E, {\bf r}_i) = - {\rm Tr} [ {\rm Im} G^r_C(E, {\bf r}_i)]/\pi$. In the top and middle panels, the LDOSs for the dispersive conduction band with $E=50$ meV (the valence band has similar results) and the flat-bands with $E=0$ in disorder-free case are plotted. For the dispersive bands, the wavefunctions extend over the entire TBG nanoflake, while for the flat-bands, the wavefunction is mostly localized within the AA-stacking region inside the TBG nanoflake. In the presence of an on-site disorder with strength $W=2.7$ eV, whose LDOSs are shown in the bottom panel in Fig.~\ref{fig.LDOS_SpectralFlow}, the localized wavefunctions spread out due to scattering from disorder and become extended. Consequently, the overlap between wavefunctions in adjacent AA-stacking regions is enlarged by disorder as illustrated in the inset, leading to enhanced inter-QD tunneling and, meanwhile, increased mobility.

To quantitatively characterize this disorder-enhanced inter-QD tunneling, we consider a unit supercell shown in the TBG nanoflake in Fig.~\ref{fig.Setup_Band}(a) with width $\mathcal{D}_W$ and length equal to the moir{\'e} period $L_m$. By imposing periodic boundary conditions along the transport direction ($x$-direction), this unit supercell can be mapped into a disordered cylinder [see the inset of Fig.~\ref{fig.LDOS_SpectralFlow} (d)]. Threading a magnetic flux 
$\phi$ through the cylinder yields the spectral flow~\cite{Tobe2007Universal, Titum2016Anomalous, Asboth2017Spectral}, namely the evolution of the bands $E_n(\phi)$ with flux. Figures~\ref{fig.LDOS_SpectralFlow} (b) and (c) show the spectral flow for the clean case ($W=0$) and a disordered case ($W=2.7$ eV), respectively. In the disordered case, valley degeneracy is lifted by disorder and cutoff edges, resulting in four low-energy bands around the CNP highlighted by the red lines in Fig.~\ref{fig.LDOS_SpectralFlow} (c), labeled $n=1$ to 4 in energy-ascending order. Since anti-crossings generally occur in the presence of disorder (see S.M.~\cite{SM}), the bands are well separated, so an effective tunneling amplitude can be defined for the lowest-energy bands:
\begin{align}
\tilde{t}_n \equiv  \frac{1}{2\pi} \int^{\pi}_{-\pi} e^{-i \phi} E_n(\phi) d \phi  
= \frac{1}{\pi} \int^{\pi}_{0} \cos{\phi} E_n(\phi) d \phi ,
\end{align}
where in deriving the second expression we have used the time-reversal symmetry: $E_n(\phi) = E_n(-\phi)$. The effective hopping $\tilde{t}_n$ characterizes the band-width of the disordered flat-bands, which in turn reflects the electron mobility of the disordered flat-bands. Figure~\ref{fig.LDOS_SpectralFlow} (d) shows the average effective inter-QD tunneling strength $\langle |\tilde{t}_n| \rangle$ of the lowest four bands, as functions of the disorder strength $W$. Here the disorder strength $W$ starts from 0.1 eV [for $W=0$ band-crossing happens so that $\tilde{t}_n$ is ill-defined (see S.M.~\cite{SM})]. For marginal disorder, the four bands remain nearly flat and the effective hopping strength is almost zero. As $W \in [0.4, 3.2]$ eV, the four bands become disentangled and all $\langle |\tilde{t}_n| \rangle$ increase, indicating the disorder-enhanced mobility of these bands. When $W>3.2$ eV, $\langle |\tilde{t}_n| \rangle$ starts decreasing. The behavior of $\langle |\tilde{t}_n| \rangle - W$ is in qualitative agreement with quantum transport results in Fig.~\ref{fig.Setup_Band} (c). In contrast, for $\theta \approx 3.15^{\circ}$, all $\langle |\tilde{t}_n| \rangle$ decrease with $W$ [see Fig.~\ref{fig.LDOS_SpectralFlow} (e)], explaining the AL observed in Fig.~\ref{fig.m10n11}. Therefore, the transport behaviors are well captured by the inter-QD tunneling picture and the disorder-enhanced conductance indeed originates from the broadening of wavefunctions.

\emph{Discussion and Conclusion.}
To conclude, we have introduced a quasi-1D TBG nanoribbon as a reliable platform for studying disorder effects on moir{\'e} flat-bands. Using a fully quantum, atomistic approach, we achieve the first quantitative calculation of conductance at the magic angle. The conductance exhibits an intriguing enhancement at moderate disorder in the flat-band of magic-angle TBG, in sharp contrast to dispersive bands and larger twist angles. This unique phenomenon arises from the disorder-driven expansion of real-space localized flat-band wavefunctions, which broadens their spatial extent and enhances inter-moir{\'e} QD tunneling, thereby accounting quantitatively for the observed transport phenomena.

For other moir{\'e} flat-band systems, wavefunctions near the magic angle also exhibit strong localization~\cite{Angeli2021Gamma, Yang2024Evolution, Naik2018Ultrafast, Zhang2020Flat, Venkateswarlu2020Electronic, Li2021Lattice}, suggesting that similar disorder-induced broadening is expected. This implies that our conclusions are not limited to magic-angle TBG but can be extended to a wider range of moir{\'e} flat-band systems. Our calculations reveal a trend of conductance delocalization at the mesoscopic scale. Although this does not clarify whether the system ultimately transitions to a metallic state or an Anderson insulator in the thermodynamic limit, the mesoscopic scale is directly relevant to current experimental samples, which are typically micrometer-sized and consist of a few hundred moir{\'e} supercells. At this experimentally accessible scale, flat-band electrons can maintain quasi-metallic behavior despite disorder, allowing these systems to be treated as good metals when investigating Coulomb-interaction effects. Taken together, our work emphasizes the unconventional role of disorder in flat-band moir{\'e} systems and provides new insights into understanding exotic transport phenomena, such as the fractional QAH effect recently observed in  MoTe$_2$~\cite{Cai2023Signatures, Park2023Observation, Xu2023Observation}.

\emph{Acknowledgements.}
This work is supported by the National Key R{$\&$}D Program of China (Grant Nos. 2024YFA1409003 and 2022YFA1403700), the National Natural Science Foundation of China (Grants Nos. 12304070, 12350401, 12404056, and 12447123), the Shanghai Science and Technology Innovation Action Plan (Grant No. 24LZ1400800), and the Natural Science Foundation of Tianjin,China (Grant No. 24JCQNJC01910).


\begin{thebibliography}{10}
\bibitem{CNPDefinition} The CNP changes if a ribbon geometry is considered. The CNP is 0.8084 eV for a 2D periodic structure while 0.783 eV for a ribbon structrue. The zero-energy point is always chosen at the CNP.
\end{thebibliography}
\end{document}


\title{Supplementary Material for ``The fate of disorder in twisted bilayer graphene near the magic angle''}

\author{
Zhe Hou$^{1,*}$,
Hailong Li$^{2}$,
Qing Yan$^{2}$,
Yu-Hang Li$^{3,4}$,
and Hua Jiang$^{2,\dagger}$
}

\affiliation{$^{1}$School of Physics and Technology, Nanjing Normal University, Nanjing 210023, China}
\affiliation{$^{2}$Interdisciplinary Center for Theoretical Physics and Information Sciences (ICTPIS), Fudan University, Shanghai 200433, China}
\affiliation{$^{3}$School of Physics, Nankai University, Tianjin 300071, China}
\affiliation{$^{4}$State Key Laboratory of Surface Physics, Fudan University, Shanghai 200433, China}

\email{zhe.hou@nnu.edu.cn}
\email{$^{\dagger}$jianghuaphy@fudan.edu.cn}

\maketitle
\tableofcontents

\newpage

\section{Transport device and band structures for TBG at twist angle $\theta \approx 3.15^{\circ} $.  } 

\begin{figure}[h]
\includegraphics[width=12cm, clip=]{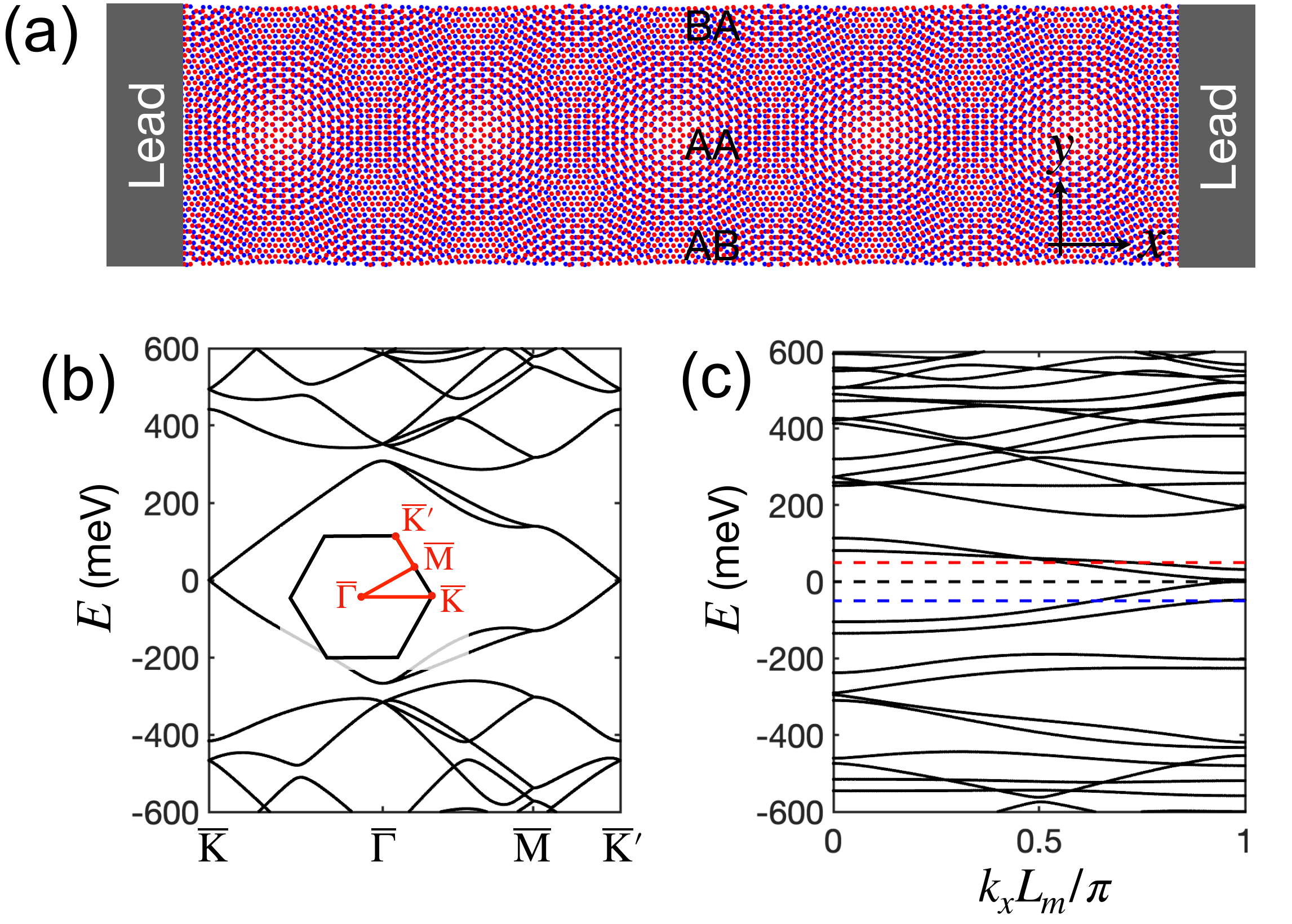}
\caption{Numerical results at twist angle $\theta \approx 3.15^{\circ}$, characterized by two commensurate numbers $m=10$ and $n=11$. (a) Schematic diagram for the transport device, showcasing the moir{\'e} pattern of TBG nanoflake with a width $N=23$ along $y$-direction in the central region. (b) Band structure of 2D TBG superlattice along $\rm \overline{K}-\overline{\Gamma}-\overline{M}-\overline{K}'$ in the moir{\'e} Brillouin zone (shown in the inset). (c) Band structure of a quasi-1D TBG nanoribbon under flat cutoff boundaries with width $N=23$. The red, black, and blue dashed lines denote the Fermi energies of $E_F=50$ meV, 0, and -50 meV, respectively, corresponding to Fig.~3 in the main text.   }
\label{fig.N23m10n11}
\end{figure}

To understand the transport results in Fig.~3 in the main text, we show the schematic diagram of the transport device, and the band structures at twist angle $\theta \approx 3.15^{\circ}$ in Fig.~\ref{fig.N23m10n11}, as determined by two commensurate parameters $m=10, n=11$. Figure~\ref{fig.N23m10n11} (b) shows the band structure of a 2D periodic TBG superlattice. The Fermi velocity at the CNP is $5.74 \times 10^5$ m/s, which is around $73\%$ of the monolayer Fermi velocity $v_F \approx 7.82 \times 10^5$ m/s. In this case the band becomes dispersive at the CNP. To preceed, we further consider a ribbon geometry with cut-off boundary terminating at AB (BA) stacking spots along $y$-direction while with periodic boundary condition along $x$-direction [see the central region of Fig.~\ref{fig.N23m10n11} (a)], whose band structure is shown in Fig.~\ref{fig.N23m10n11} (c). We see that, the band deforms due to the finite-size effect, and the valley degeneracy is lifted due to the cut-off boundaries that mixes the two valleys. Nevertheless, the dispersive feature of bands around the CNP is well maintained. Note that the CNP shifts from 0.801 eV to 0.882 eV when the geometry of TBG changes from 2D infinitely large system to a quasi-1D nanoribbon. We have set the CNP as the zero-energy point throughout this Supplementary Material. 

In Fig.~3 in the main text, the Fermi energies are fixed at the CNP of the nanoribbon, which always intersects with the dispersive bands. The average conductance $\langle G \rangle$ decreases sharply with the disorder strength $W$, which restores the AL. This behavior is different from the conductance curve at the magic-angle in Fig.~1(c) in the main text.

\section{Transport results at twist angles $\theta \approx 1.1607 ^{\circ}$ and $\theta \approx 1.0845^{\circ} $.} 
To show that the disorder-enhanced transport behavior is not a unique phenomenon for twist angle $\theta \approx 1.12^{\circ}$ (decided by the two commensurate integers $m=29, n=30$), we consider two other sets of commensurate integers: $m=28, n=29$ and $m=30, n=31$, corresponding to twist angles $1.1607^{\circ}$ and $1.0845^{\circ}$, respectively. As shown below, we find that the same transport behavior appears as long as isolated flat-bands exist. 

\begin{figure}
\includegraphics[width=12cm, clip=]{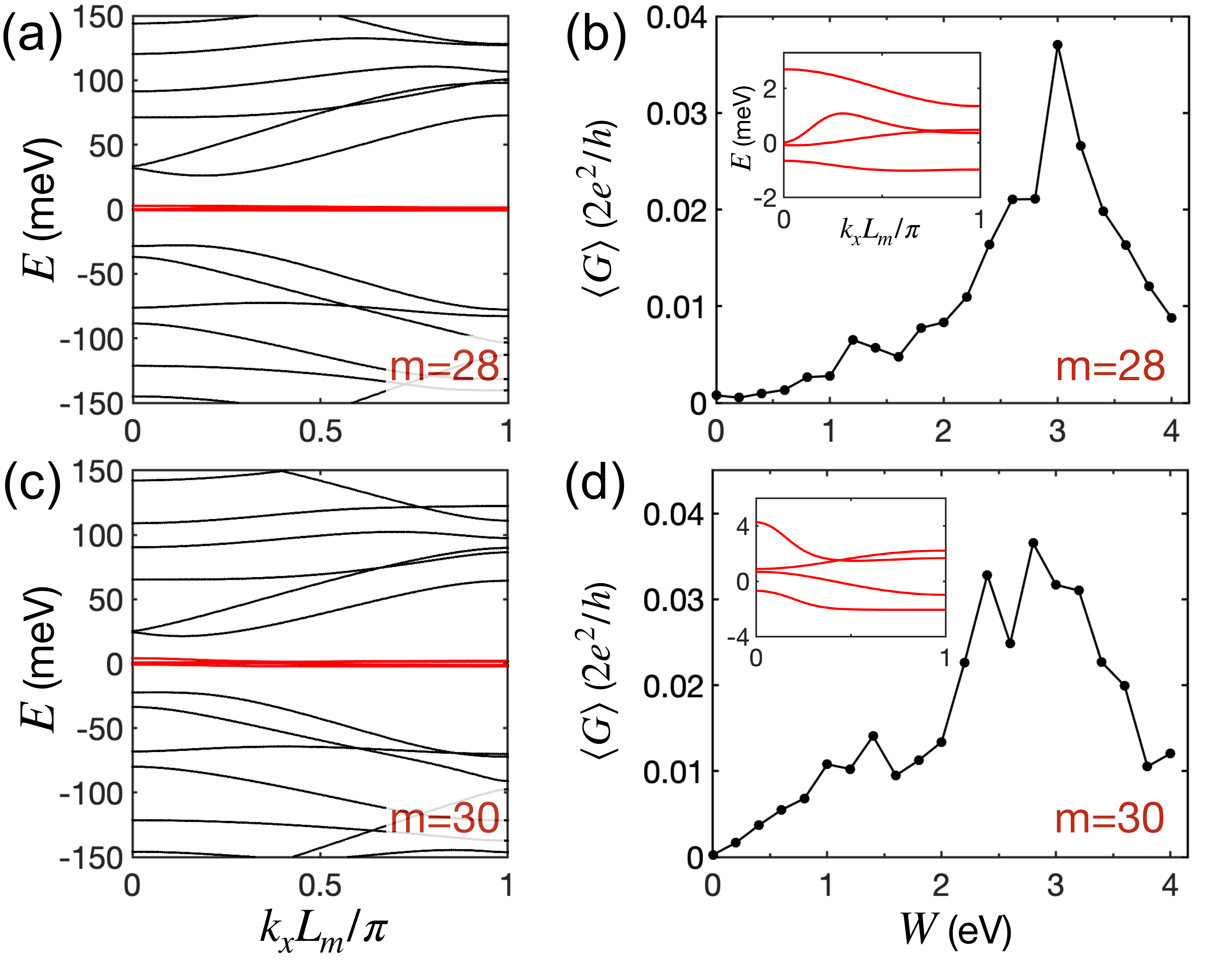}
\caption{(a) and (b): Band structure of the quasi-1D TBG nanoribbon, and the average-conductance $\langle G \rangle $ at $E_F=0$ versus the disorder strength $W$, respectively. Here the commensurate numbers are $m=28$, and $n=29$, corresponding to the twist angle $\theta \approx 1.1607^{\circ}$. The inset panel in (b) shows the zoom-in of the flat-bands in (a), as marked with red.  For (c) and (d), the results are for $m=30$, and $n=31$, corresponding to twist angle $\theta \approx 1.0845 ^{\circ}$. The band structures in (a) and (c) are obtained at $N=55$ along $y$-direction, while the conductance in (b) and (d) is calculated with $N=55$, $L=261$ under 100 disorder configurations average.}
\label{fig.m28m30}
\end{figure}

In Figs.~\ref{fig.m28m30} (a) and (c), we plot the band structures of the quasi-1D TBG nanoribbon with $(m=28, n=29)$ and $(m=30, n=31)$, respectively. Here the width of the nanoribbons is set to $N=55$ so that the nanoribbon terminates at AB/BA-stacking spots along $y$-direction. We see that there are four flat-bands (marked in red) at the CNP for each figure, which are disentangled with other remote dispersive bands with band gaps larger than 20 meV. The zoom-in of these flat-bands are plotted in the inset panels in Figs.~\ref{fig.m28m30} (b) and (d). The Fermi energy is finely tuned at the flat-bands ($E_F = 0$), and the disorder-averaged conductance $\langle G \rangle $ versus the disorder strength $W$ curves are plotted in Figs.~\ref{fig.m28m30} (b) and (d). When $W< 3 $ eV, a prominent conductance enhancement is obtained for both $(m=28, n=29)$ and $(m=30, n=31)$. If $W>3 $ eV, the conductance decreases as $W$ increases, in consistent with the conventional AL. These results agree quanlitatively with the case of $(m=29, n=30)$ in the main text, reaffirming the universal disorder-enhanced conductance behavior as long as isolated flat-bands exist.

\section{Disorder-dependent spectral flow of the TBG cylinder} 
\begin{figure}
\includegraphics[width=15cm, clip=]{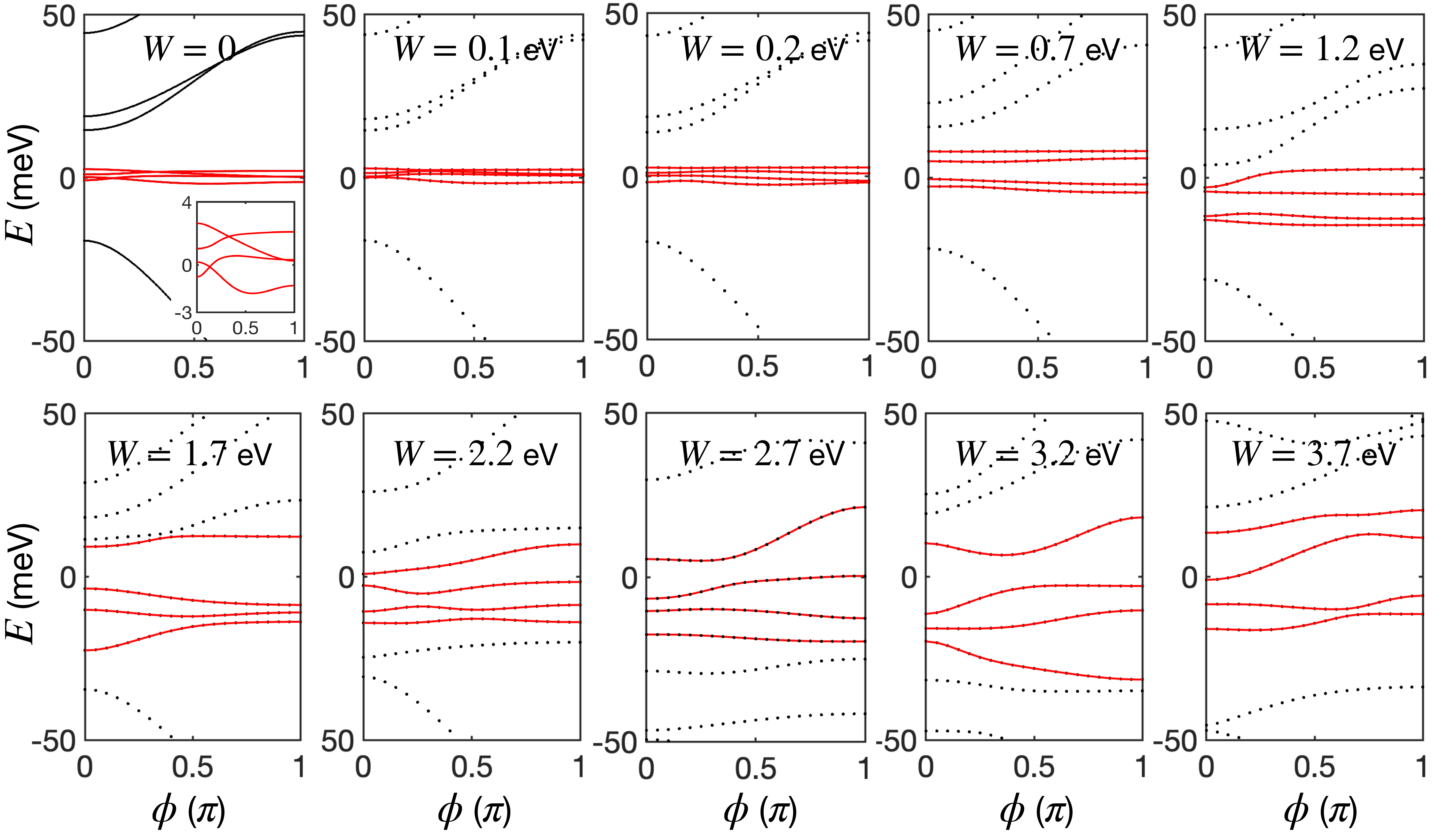}
\caption{Spectral flow of a disordered cylinder of a TBG nanoflake with periodic boundary condition, whose length is the moir{\'e} period $L_m$ and width is $N=55$. The twist angle is $\theta \approx 1.12^{\circ}$ characterized by two commensurate numbers $m=29$ and $n=30$. From the left to the right the disorder strength $W$ is increased from zero to 3.7 eV. The inset in $W=0$ panel shows the zoom-in of the four flat-bands, which are marked in red color. Only one configuration is adopted for each disorder strength.  }
\label{fig.SpectrumFlowm29n30}
\end{figure}

\begin{figure}
\includegraphics[width=15cm, clip=]{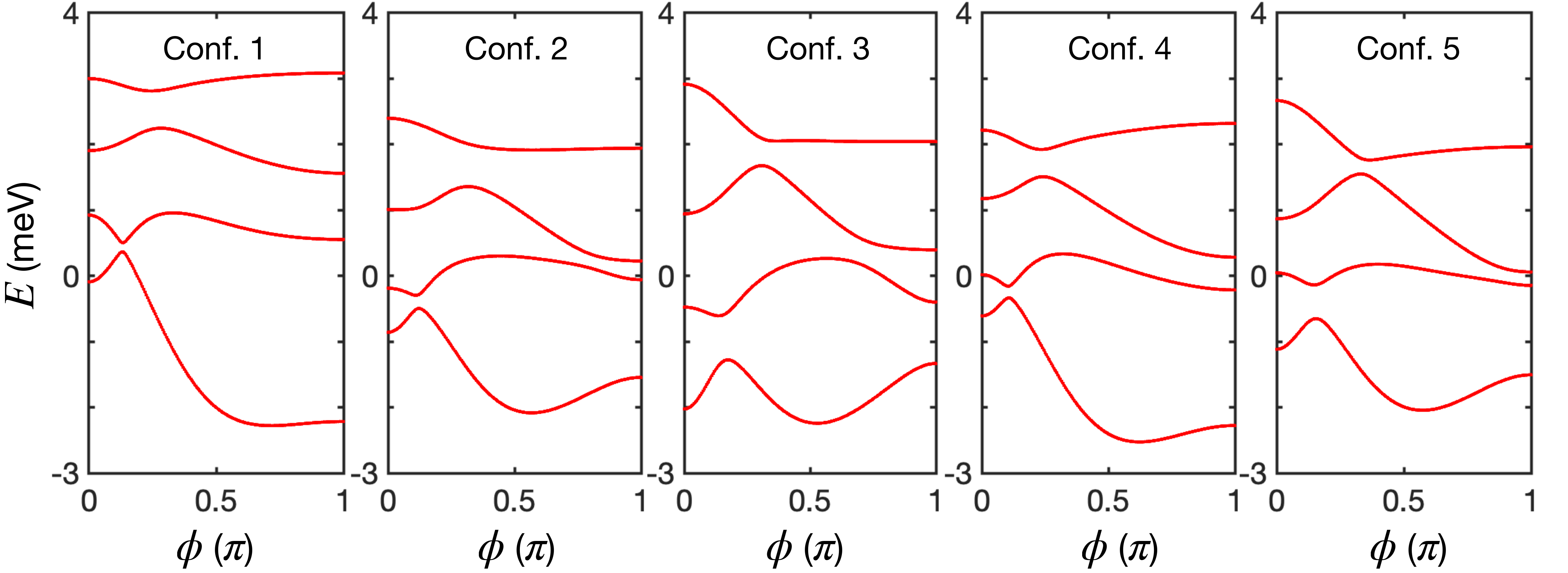}
\caption{Spectral flow of a disordered TBG cylinder. The disorder strength is $W=0.1$ eV. Five different disorder configurations are adopted here. Other parameters are the same as in Fig.~\ref{fig.SpectrumFlowm29n30}.  }
\label{fig.BandAnticrossing}
\end{figure}

To comprehensively understand the disorder effect on the mobility of TBG, in Fig.~\ref{fig.SpectrumFlowm29n30}, we plot the spectral flow of a disordered TBG cylinder under different disorder strength $W$. Here, one configuration is considered for each disorder strength. For $W=0$ (disorder-free system), four isolated flat-bands (marked in red) gapped from the remote dispersive bands appear near the CNP. These flat-bands are entangled with each other due to band crossing (inset). In the presence of weak disorder, band crossing is eliminated and the four flat-bands become disentangled from each other. This is further confirmed by the spectral flow from five random configurations at $W=0.1$ eV shown in Fig.~\ref{fig.BandAnticrossing}, where all configurations exhibit sizable band gaps near the band crossing point. As $W$ increases, this disentanglement becomes stronger and the flat-bands are pushed away from each other. Simultaneously, the flat-bands become more dispersive, with their band-width becoming larger. This demonstrates the enhanced mobility, and alternatively, the increased effective inter-QD tunneling strength $\langle | \tilde{t}_n | \rangle $ for flat-band electrons in Fig.~4(d) in the main text.

